Stresa, Italy, 25-27 April 2007# MEMS Q-FACTOR ENHANCEMENT USING PARAMETRIC AMPLIFICATION: THEORETICAL STUDY AND DESIGN OF A PARAMETRIC DEVICE

*Laetitia Grasser, Hervé Mathias, Fabien Parrain, Xavier Le Roux and Jean-Paul Gilles*
Institut d'Electronique Fondamentale UMR 8622 - Université Paris-Sud - France**ABSTRACT**

Parametric amplification is an interesting way of artificially increasing a MEMS Quality factor and could be helpful in many kinds of applications. This paper presents a theoretical study of this principle, based on Matlab/Simulink® simulations, and proposes design guidelines for parametric structures. A new device designed with this approach is presented together with the corresponding FEM simulation results.## 1. INTRODUCTION

Quality Factor (Q) is one of the most important characteristics of MEMS resonators, especially if they are used to build sensors based on frequency monitoring. The corresponding frequency resolution, and thus the system's sensitivity, is then indeed directly linked to Q. The higher the value of Q, the higher the micro-system's performance. These MEMS resonators are indeed found in many applications were a high sensitivity is needed: inertial sensors, mass sensors … To get high Q-values, these micro-systems generally rely on the use of vacuum packaging, air damping being an important limitation to the quality factor [1].

Another possibility to get high Q-values could be to externally increase the quality factor. An interesting technique to artificially improve the quality factor is called parametric amplification and consists in modulating the structure's stiffness at a harmonic frequency of the device's resonant frequency.

This modulation results in an increase of the oscillation amplitude at the device's resonant frequency and thus an increase of Q. Quality factor improvement by a factor of more than 10 have been reported in the case of a thermal stiffness modulation obtained using a pulsed laser locally heating the device [2] or by using electrostatic modulation exciting two different modes on a torsional microresonator [3].

Parametric amplification using MEMS is also encountered in other applications than Q-factor enhancement: in [4], MEMS pumped capacitors are used to constitute an electrical parametric amplifier, on the same principle as those developed in the 1960's. In [5], the cubic non linearities and their impact on the stability of the parametric resonance are exploited to build a femtogram resolution mass sensor.

Our aim is to develop an easy to implement parametric device with better Q-factor enhancement possibilities.

This paper first presents a theoretical high level study of the damped Mathieu equation, that governs the considered parametric devices and is shown on Figure 1, to determine the most critical parameters and have an insight of system constraints for different potential applications.

In the second part of this paper, we use this theoretical study to design a new MEMS structure allowing an integrated implementation of the parametric amplification.

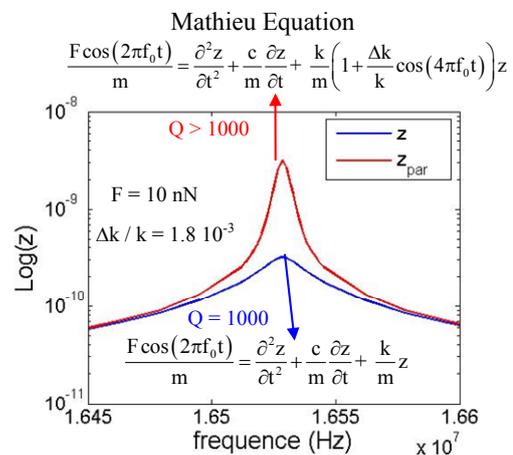

*Figure 1: Vibration amplitude increase at resonant frequency using parametric amplification*

The proposed parametric amplification principle is then validated and the corresponding characteristics are estimated using FEM simulations.

## 2. SYSTEM-LEVEL STUDY OF PARAMETRIC AMPLIFICATION

In order to be able to use parametric amplification at a system level for Q enhancement, we found it interesting to study the role and the criticality of the different parameters appearing in the damped Mathieu equation. The MEMS device considered is modeled as a 2nd order low-pass filter with initial quality factor Q and resonant frequency $f_0$. It is put into forced vibrations by an AC actuation force, with a small enough amplitude that guarantees that the resonator's behavior remains linear even after parametric amplification. The Mathieu equation is rewritten below in order to clearly show the considered parameters: the phase difference between the modulation signal and the actuation signal (Φ), the actuation and modulation frequencies (resp. $f_A$ and $f_P$), the relative stiffness modulation magnitude ($\Delta k/k$) and the initial value of Q.

©EDA Publishing/DTIP 2007       ISBN: 978-2-35500-000-3

$$\ddot{z} + \frac{\omega_0}{Q}\dot{z} + \omega_0^2(1 + \frac{\Delta k}{k}\cos\omega_P t)z = \frac{F_0 \cos(\omega_A t + \phi)}{m} \quad (1)$$

We have performed Matlab/Simulink® simulations to determine the impact of all these parameters, with only one parameter varying at a time. The default configuration for the non-varying parameters was as follows: Q=1000, $f_A = f_0$, $f_P = 2*f_A$, Φ=0 and $\Delta k/k$=0.18% like on figure 1. With these parameters values, an amplitude gain of 10 is obtained at the resonant frequency thanks to parametric stiffness modulation.

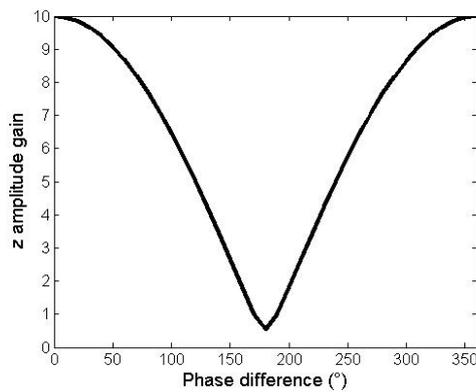

*Figure 2: Impact of phase difference between actuation and modulation signals on parametric amplification*

### 2.1. Impact of phase difference

We have first studied how the obtained displacement gain is affected by a non zero phase difference between actuation and modulation signals. As shown on figure 2, a phase difference of up to 20° has not much impact on the obtained amplitude gain. However, if the phase difference is too large, the parametric amplification is less efficient. With a phase difference near 180°, the vibration amplitude at the resonant frequency is even lower than without parametric modulation. However, the performed simulations have shown that the phase control is not critical.

### 2.2. Impact of actuation frequency choice

The second set of performed simulations was aimed at determining the influence on the obtained amplification of any discrepancy between the actuation frequency and the resonant frequency. For these simulations, the modulation signal has been kept at twice the actuation signal frequency. As shown on figure 3, a slight difference has an important negative impact on the obtained amplification. A closed-loop configuration to actuate the resonator would thus probably be the best choice to perform parametric amplification.

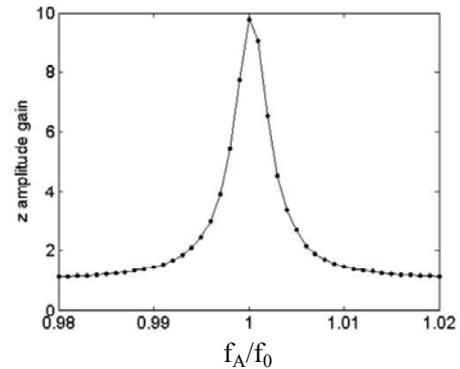

*Figure 3: Impact on parametric amplification of actuation frequency choice with respect to resonant frequency*

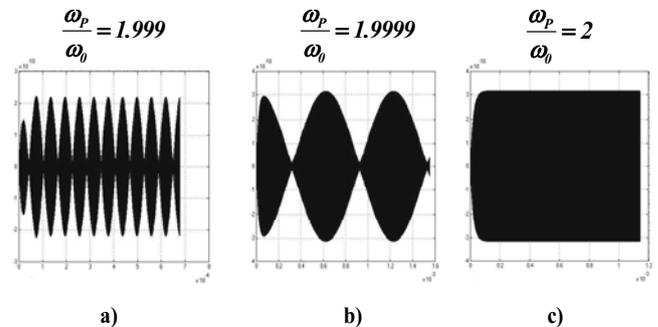

a)          b)          c)

*Figure 4: Obtained amplified responses with varying ratios between actuation and modulation frequencies*

### 2.3. Impact of modulation frequency choice

The frequency ratio between the modulation and the actuation signals has to be exactly equal to two. As shown on figure 4, the slightest difference causes an amplitude modulation on the obtained response. The nearer the ratio to two, the lower the amplitude modulation frequency. The observed movement amplification also depends on the ratio between the modulation and actuation frequencies. It is maximal for a ratio of exactly two. The modulation signal should thus be generated from the square of the actuation signal to guarantee the ratio between modulation and actuation frequencies

### 2.4. Impact of stiffness modulation amplitude

Looking at the impact of the relative stiffness modulation amplitude, one can see, as shown on figure 5, that nearly no parametric amplification is obtained for a large set of values. Once the amplification becomes large enough (>

10), it then becomes very sensitive to relative stiffness variations. To accurately control the amplification gain, and so the Q factor gain, the control on the stiffness has to be very thorough. The way that the stiffness modulation is induced has thus to be chosen adequately during the parametric structure design. One must be able to create a sufficient stiffness variation, but not too large to keep the structure from breaking or reaching an amplification level controlled by the non linearities of the device.

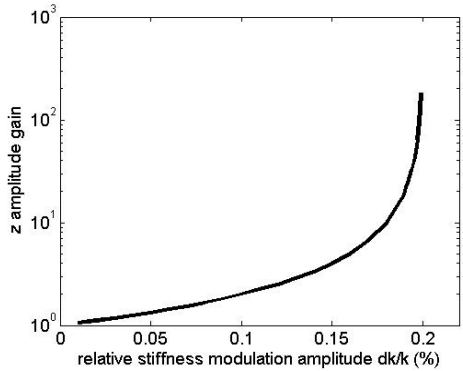

*Figure 5: Parametric amplification variation with respect to relative modulation amplitude*

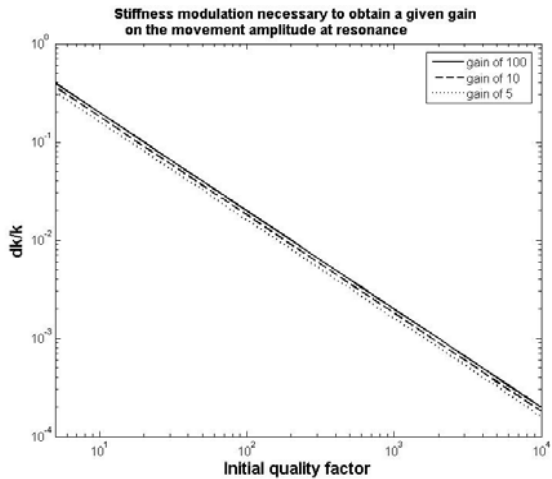

*Figure 6: Impact of initial quality factor on necessary stiffness modulation for a given parametric gain*

### 2.5. Impact of stiffness modulation amplitude

Finally, as shown on Figure 6, for a given amplitude gain, the necessary relative stiffness modulation magnitude depends on the initial value of Q. The lower the latter, the larger the former has to be. For an initial Q value of 10, a 20% stiffness modulation is necessary to get a gain of 10 on the oscillation amplitude at resonant frequency whereas a modulation of only 0.2% is sufficient for the same gain if the initial Q value is 1000. This shows that the parametric amplification may more easily be used to enhance the performance of vacuum packaged resonators, that generally have a high Q value, than to replace altogether vacuum packaging for high sensitivity sensors. In the latter case, the stiffness modulation to be performed would be too large to be realistic or, if it is feasible, would probably have negative consequences on the initial Q-factor value.

This theoretical study has allowed us to better understand how the parametric amplification works and has given us some guidelines to design parametric structures that will be exploited in the next part.

### 3. DESIGN OF A NEW PARAMETRIC DEVICE

As mentioned in the introduction, in existing parametric structures, the stiffness modulation is obtained either thermally or electrostatically. If the first solution offers a good decoupling between actuation and stiffness modulation, it has nonetheless some drawbacks with respect to what we have seen in the precedent part: there is an inherent latency of the device to the thermal effect that may limit the control that we have on the Q-factor gain in a closed loop system. Thermal stiffness modulation also necessitates an extra power consumption that may be prohibitive in some embedded micro-systems. Electrostatic solutions seem to be a better choice. However, in existing electrostatic parametric structures, the parametric amplification is induced using the non-linearities in the electrostatic actuation. There is, in this case, a strong coupling between the actuation signal and the stiffness modulation that could again impede the control on the Q-factor gain.

Our idea was to find a way to electrostatically modify the device's stiffness without affecting the actuation scheme. In our proposed parametric device, the stiffness modulation is performed perpendicularly to the oscillation direction, which allows a greater decoupling between excitation and modulation. Figure 7 shows the ANSYS® model used to optimize the design.

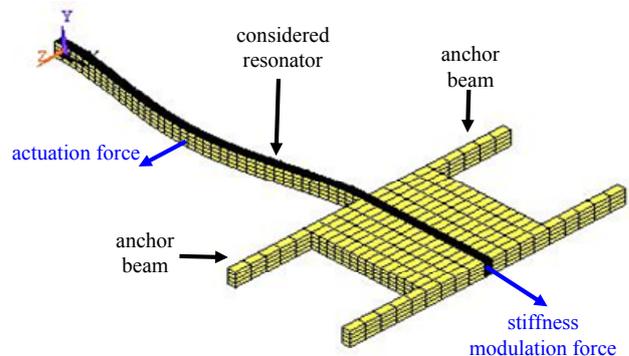

*Figure 7: Proposed parametric device ANSYS model*

*Laetitia Grasser, Hervé Mathias, Fabien Parrain, Xavier Le Roux and Jean-Paul Gilles*
*MEMS Q-FACTOR ENHANCEMENT USING PARAMETRIC AMPLIFICATION*

The considered resonator is a 200µm long, 5µm wide and 10µm thick Si beam that is attached on one side to a fixed pad and on the other side to a mass only able to move along the x-axis. By applying a force on the latter, the considered resonator's stiffness will be modified. The applied modulation force is composed of two terms: a static force fx_stat and a varying force fx_var at twice the resonant frequency.

Static simulations have been performed to dimension the anchor beams and the moving mass so that the actuation force only induces a negligible movement in the z-direction at the anchor point between the resonator and the moving mass. To simplify the remaining ANSYS® analyses, the model used was a single beam with an end fully clamped and the other only clamped in the z-direction. The actuation force is applied in the middle of the beam and the stiffness modulation force on the partially clamped end.

Modal simulations have been performed to quantify the impact of the static part of the modulation force on the resonator's resonant frequency and deduce the corresponding stiffness variation. If the resonant frequency is given by

$$f_0 = \frac{1}{2\pi}\sqrt{\frac{k}{m}} \quad (2)$$

as in the classical spring-mass-damper model used for MEMS resonators, the link between the relative resonant frequency and the stiffness variation is given by

$$\frac{\Delta f}{f_0} = \frac{1}{2}\frac{\Delta k}{k} \quad (3)$$

To validate the precedent theoretical study, an initial quality factor value of 40 has been imposed. In order to get a gain of 100 on the amplitude at resonance, we need to apply a relative stiffness modulation of the order of 5%. Modal simulations have shown that such a variation may be obtained using fx_stat = 2.47 $10^{-3}$ N and fx_var = $10^{-3}$ N. The corresponding results are shown on figure 8.

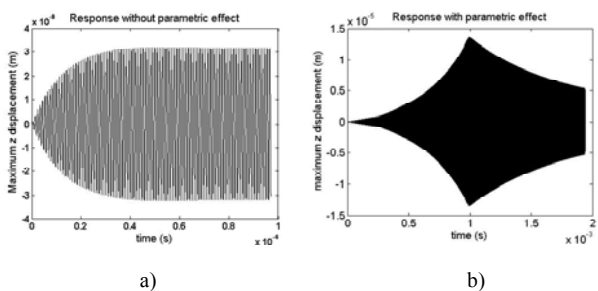

*Figure 8: Transient response of the new parametric device without a) and with b) parametric amplification*

Without parametric amplification (fx_var = 0), an amplitude of 30nm is observed at resonant frequency (Figure 8 a)). When fx_var = $10^{-3}$ N, a parametric amplification is effectively observed, validating the stiffness modulation principle (Figure 8 b)). However, there is an unexpected transient phase with a maximum amplitude gain of about 500. Apparently, the applied stiffness modulation was larger than needed and the corresponding parametric amplification was thus much higher, as expected from part 2.4. The non-linearities in the structure, that were taken into account during the simulation, limit the obtained amplification during the transient phase. The steady-state gain seems to be more of the order of 100, if steady state is effectively reached a few milliseconds after the end of our simulation. This would be coherent with what was expected. But since each transient simulation took more than a day using a 2.6 GHz Intel Pentium processor, we could not check if this is the case. Physical characterizations will be necessary to find out the effective gain on the Q-factor.

In order to implement the described parametric amplification principle, we propose to use a device similar to the one shown on figure 9.

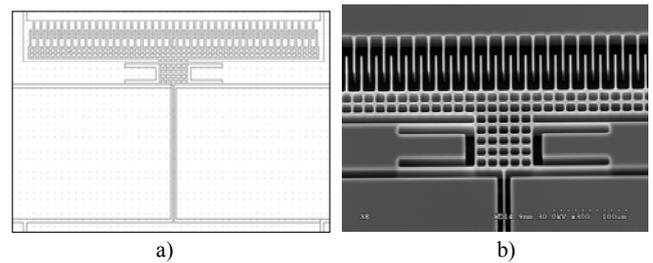

a)     b)

*Figure 9: Layout a) and MEB photograph b) of proposed parametric structure*

A comb drive is used to apply the stiffness modulation force and electrodes along the resonator allow actuating and sensing. However, a force of $10^{-3}$ N, like in the situation considered in the precedent simulations, is clearly not achievable: the electrostatic force created by a comb drive is given by

$$F_{cd} = \frac{1}{2}\frac{n\,t\,\varepsilon_0\,V^2}{g} \quad (4)$$

where n is the number of comb fingers, t is the thickness of the comb fingers, g the gap between them and V the applied voltage between the electrodes. A realistic choice of parameters would be n= 70, t/g = 20, V = 40 V. For such a configuration, the obtained electrostatic force would be $F_{cd} \approx 10\,10^{-6}$ N, that is two orders of magnitude lower. If not adapted for Q-factor enhancement for resonators working at atmospheric pressure, this device could be used in the case of vacuum packaged structure.

*Laetitia Grasser, Hervé Mathias, Fabien Parrain, Xavier Le Roux and Jean-Paul Gilles*
*MEMS Q-FACTOR ENHANCEMENT USING PARAMETRIC AMPLIFICATION*

The achievable stiffness modulation with our considered resonator would be of 0.03%. This could allow performing parametric amplification on devices with an initial Q of at least 5000 (Figure 6), which is common place for resonators working at low pressures.

The resonators dimensions could also be adapted so that the comb drive could create a higher relative stiffness modulation. If only the length of our resonator is increased, the maximum relative stiffness modulation induced by the comb drive will be larger: for example, for a 400µm long resonator with the same thickness and width as before, the achievable stiffness modulation would be 0.115%. This allows applying parametric amplification for devices with initial Q values around 900.

Depending on the material used and the ambient pressure at which the device works, an optimized parametric structure using our proposed principle should most of the time be useable.

## 4. CONCLUSION

A theoretical study of the parametric amplification as a way to enhance the Q-factor of MEMS resonators has been conducted. The most critical parameters have been identified and some guidelines have been deduced for the design of parametric devices. The initial quality factor of the structure has to be large enough so that the necessary stiffness modulation to obtain the desired Q-factor gain may be effectively implemented. The stiffness modulation has also to be precisely controlled. Following these guidelines, a new and easily implemented parametric device structure has been proposed and studied. FEM simulations have validated the proposed parametric amplification principle.

Physical characterization will nonetheless have to be performed to confirm these results. The proposed device's dimensions could be easily adapted so that it could be used in many applications, as long as the device's initial quality factor is large enough.